\documentclass[a4,center,fleqn,twocolumn]{article}
\usepackage[utf8]{inputenc}
\usepackage{color}
\usepackage{graphicx}
\usepackage{subfigure}
\usepackage{amsmath}
\usepackage{url}
\usepackage{hyperref}

\usepackage{layouts}
\usepackage[authoryear,round]{natbib}

\begin{document}


\title{ProQ3D: Improved model quality assessments using  Deep
  Learning.}  
\author{Karolis
  Uziela\,$^{1\dagger}$, David Men\'{e}ndez
  Hurtado\,$^{1\dagger}$, Bj\"orn Wallner\,$^{2}$ and
  Arne Elofsson\,$^{1*}$ \\
  \small $^{1}$ Department of Biochemistry and Biophysics and Science for Life Laboratory, Stockholm University, 171 21 Solna, Sweden and \\
  \small $^{2}$ Department of Physics, Chemistry and Biology (IFM) /Bioinformatics. Link\"oping University, 581 83 Link\"oping, Sweden. \\
  \small $^\ast$To whom correspondence should be addressed.
  \small $^{\dagger}$Contributed equally
  }

\maketitle


\begin{abstract}
\textbf{Summary:} Protein quality assessment is a
  long-standing problem in bioinformatics. For more than a decade we have
  developed state-of-art predictors by carefully selecting and 
  optimising inputs to a machine learning method. The
  correlation has increased from 0.60 in ProQ to 0.81 in ProQ2
  and 0.85 in ProQ3 mainly by adding a large set of carefully tuned
  descriptions of a protein. Here, we show that a substantial improvement can
  be obtained using exactly the same inputs as in ProQ2 or ProQ3 but
  replacing the support vector machine by a deep neural network. This
  improves the Pearson correlation to 0.90 (0.85 using ProQ2
  input features). \\
  \textbf{Availability:} ProQ3D is freely available both as a
  webserver
  and a stand-alone program at http://proq3.bioinfo.se/ \\
  \textbf{Contact:} \href{arne@bioinfo.se}{arne@bioinfo.se}\\
\textbf{Supplementary information:} Supplementary data are available
at \textit{arXiv} online.
\end{abstract}

\section{Introduction}

In 2003 we developed the first real model quality estimation program
ProQ~\citep{Wallner12717029}. In contrast to earlier methods, such as~\cite{Park8627632}, ProQ is
not trained to recognise the native structure but to estimate the
quality of a model. ProQ uses a machine learning approach and many
features describing a protein model. In ProQ the quality is
calculated for the entire model but in 2005 we extended it to
estimate the quality of each residue~\citep{Wallner16522791}. The quality of the entire model
was then estimated by summing up the predicted qualities for each
residue. In ProQ2 profile weights were added to improve the
predictions~\citep{Ray22963006} and in ProQ3~\citep{Uziela27698390} we added
energy terms calculated from Rosetta~\citep{Leaver-Fay21187238}. The
ProQ methods have since their introduction been the best single-model based
quality assessors in CASP~\citep{Kryshtafovych26344049}.

ProQ, ProQ2 and ProQ3 use a large number of carefully tuned inputs
that are calculated from each protein model. All parameters are
optimised to be independent of protein size and to have a limited
range. These parameters are then used to train a support vector
machine using a linear kernel (ProQ used a neural network). More advanced kernels are
computationally expensive and do not produce any
significant improvements. This means that ProQ2 and ProQ3 basically are
linear combinations of a large set of features that all independently
show a weak correlation with model quality. When these features are
combined a much better correlation is achieved. However, the ProQ2 and ProQ3
methods can not identify relationships where the different features provide
opposite results, i.e. it can not identify more complicated,
non-linear, relationships between the features.

In the last few years machine learning using so called deep neural
networks has proven to be clearly superior to other machine learning
methods. These networks are able to identify non-linear relationships
between input features. We find that using identical inputs as in
ProQ2 and ProQ3 but replacing the support vector machine with a deep
neural network a substantial improvement can be obtained for both
ProQ2 and ProQ3. The improvement is of a similar magnitude as obtained
by the years of optimisation that was used to optimize the input
features for ProQ3, and the gap to the consensus based assessor,
Pcons~\citep{Lundstrom11604541} has never been this small (CC=0.90 vs
0.95).



\section{Methods}

As in ProQ2 and ProQ3 a
large number of features are calculated describing a model and then
used to predict the quality, as measured by the
S-score~\citep{Ray22963006}, for a single residue.  Training was done
using all models from CASP9 and CASP10, this is substantially more
than we could use when training ProQ3.  Testing
was done on all models from CASP11 excluding cancelled targets and
targets shorter than 50 residues.
The Pearson correlation for local and global quality was used to evaluate the
performance. 

The learning was performed using the Keras Python library with the
Theano backend. We used two dense hidden layers with 200 and 600
neurons respectively. Increasing the number of layers and neurons did
not improve the results.  The final model was trained with Adadelta
and $10^{-11}$ penalty for the $L^2$ regularization and shuffling the
training data, for details see supporting information.

\section{Results and Discussion}

\begin{table}[t]
\caption{
{\bf Performance of different QA methods on CASP11.  }}
\begin{tabular}{p{1.2cm}|p{1.2cm}p{1.2cm}p{1.2cm}p{1.2cm}p{1.2cm}} 
	& CC-glob & CC-target & CC-loc & CC-model & GDT\_loss \\
\hline	
  ProQ & 0.60 & 0.44 & 0.50 & 0.39 & 0.06 \\
  Qprob  & 0.71 & 0.56 & - & - & 0.07 \\
  Qmean  & 0.73 & 0.57 & 0.57 & 0.42 & 0.08 \\
  ProQ2 & 0.81 & 0.65 & 0.69 & 0.47 & 0.06 \\
  ProQ3 & 0.85 & 0.65 & 0.73 & 0.51 & 0.06 \\
  ProQ2D & 0.85 & 0.68 & 0.72 & 0.49 & 0.05\\
  ProQ3D  & 0.90 & 0.71 & 0.77 & 0.54 & 0.06 \\
\hline
 & \multicolumn{5}{c}{Consensus based methods} \\
\hline
  Pcons  & 0.95 & 0.77 & 0.87 & 0.68 & 0.07 \\
\end{tabular}
\label{tab:correlations}
\end{table}

\begin{figure}[t!]
  \includegraphics[clip, trim=0cm 0cm 0cm 1cm, width=\columnwidth]{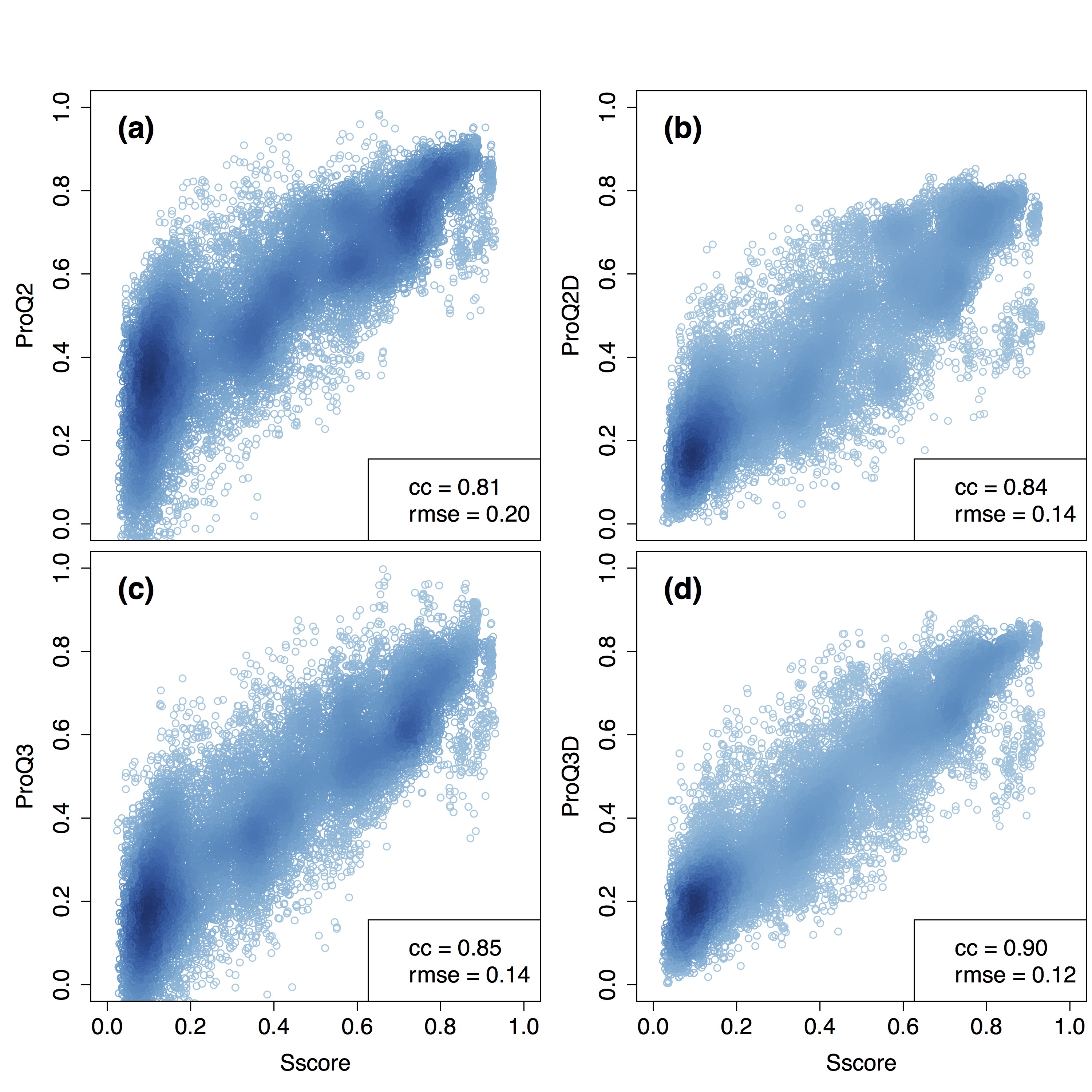}
  \caption{Predicted global model quality on CASP11 data set for ProQ2
    (top) and ProQ3 (bottom) trained with a SVM (left) or Theano
    (right). RMSE is a root mean square error between the prediction
    and the target value (lower values are better). (a) ProQ2, SVM,
    (b) ProQ2D, Theano, (c) ProQ3, SVM, (d) ProQ3D, Theano }
\label{fig:global}
\end{figure}

To estimate the global quality of a protein model with ProQ the predicted
qualities for each residue are summed and the sum is divided by the protein sequence length. The correlation of overall
(global) quality is substantially improved using either ProQ2 or ProQ3
inputs both when we calculate the correlation for all models together (CC-glob) or when the per target
correlations are studied (CC-target), Table~\ref{tab:correlations} and Figure~\ref{fig:global}. 
In addition, protein model quality assessment can be used to identify good and bad
regions of a model, i.e. the local quality. Here we obtain a similar improvement as achieved for
the global quality estimation independently if it is measured for all
residues (CC-loc) or per model (CC-model)
Table~\ref{tab:correlations}.

Finally, we examined if better top-ranked models could be selected
using the new quality estimators. We calculated the average GDT loss of the first ranked models for each method. Unfortunately, the selection of top ranked models does not show any
significant improvement between any of the top QA methods. However,
the same is observed when using Pcons, although the correlation is
probably quite close to the theoretical limit, as it is higher than the correlation between different methods used to evaluate the quality of a
model~\cite{Wallner17894353}. This indicates that to advance further
it will be necessary to use a different approach.

%
%


\section*{Funding}

This work was supported by grants from the Swedish Research Council
(VR-NT 2012-5046 to AE and 2012-5270 to BW) and Swedish e-Science
Research Center (BW). Computational resources were provided by the Swedish National
Infrastructure for Computing (SNIC) at NSC.


%
%


\bibliographystyle{genres}

\end{document}


\title{Supporting material for: "ProQ3D: Improved model quality assessments using  Deep Learning."}
\author{Karolis Uziela, David Menendez Hurtado, Nanjiang Shu, Bj\"orn Wallner and Arne Elofsson}
\date{}
\maketitle

\beginsupplement

\section{Supplementary Results}

\begin{figure}[!ht]
  \includegraphics[width=\textwidth]{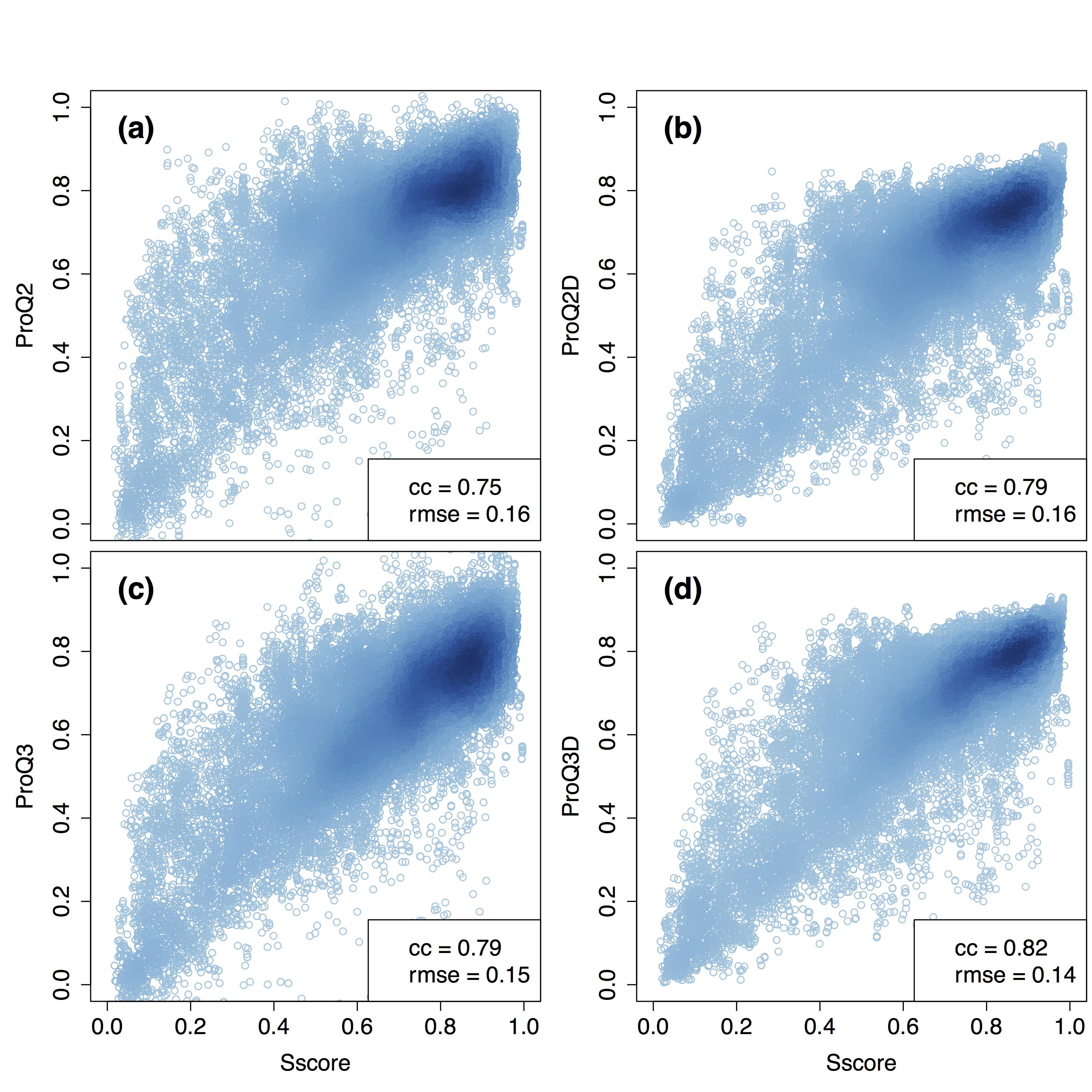}
\caption{Predicted global model quality on CAMEO data set for ProQ2 (top) and ProQ3 (bottom) trained
  with a SVM (left)   or Theano (right). RMSE is a root mean square error between the prediction and the target value (lower values are better). (a) ProQ2, SVM, (b) ProQ2D, Theano, (c) ProQ3, SVM, (d) ProQ3D, Theano}
\label{fig:global_cameo}
\end{figure}

Here, we provide results that are analogical to the ones in the main
text but on another data set - CAMEO~\citep{Haas23624946}. All CAMEO
models from 2014-06-06 to 2015-05-30 were
used. Table~\ref{tab:correlations} shows the correlations and
GDT\_loss for CAMEO data set and Figure~\ref{fig:global_cameo} shows
the model quality prediction scatter plots. As we already noted for 
CASP11, the deep learning versions of the predictors (ProQ2D and
ProQ3D) are superior to their SVM-based counterparts (ProQ2 and
ProQ3).

\begin{figure}[!ht]
  \includegraphics[width=\textwidth]{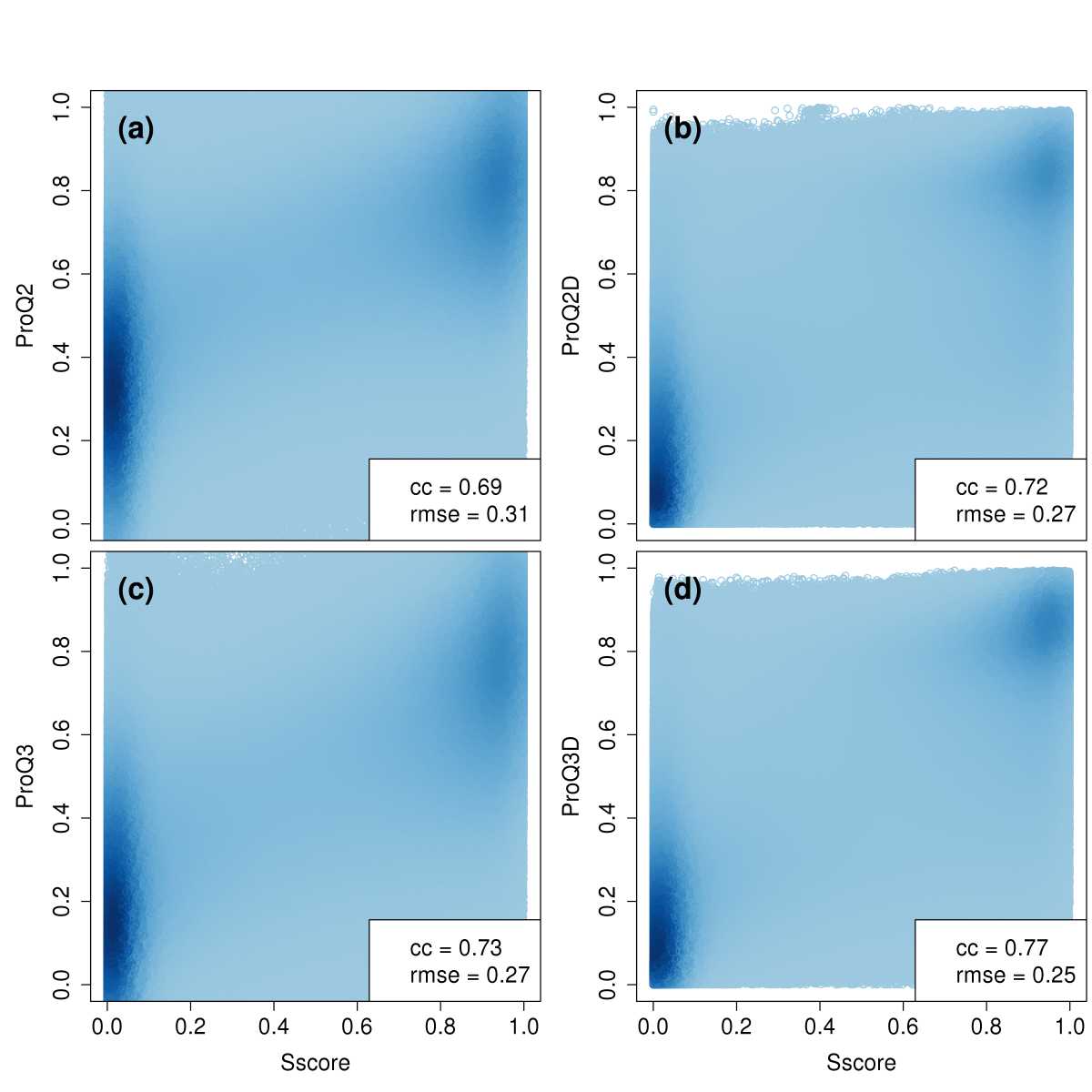}
\caption{Predicted local model quality on CASP11 data set for ProQ2 (top) and ProQ3 (bottom) trained
  with a SVM (left)   or Theano (right). RMSE is a root mean square error between the prediction and the target value (lower values are better). (a) ProQ2, SVM, (b) ProQ2D, Theano, (c) ProQ3, SVM, (d) ProQ3D, Theano}
\label{fig:local_casp11}
\end{figure}

\begin{figure}[!ht]
  \includegraphics[width=\textwidth]{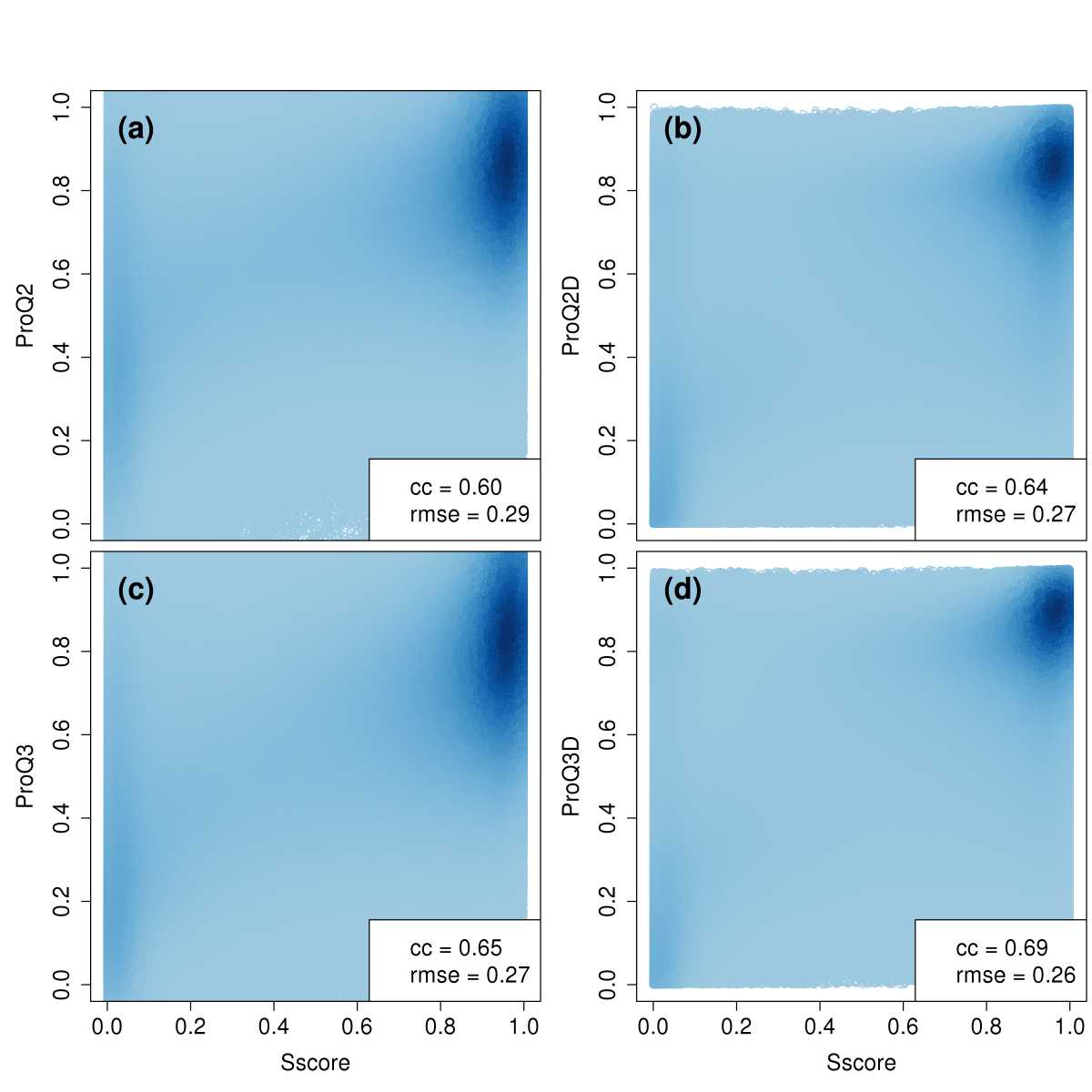}
\caption{Predicted local model quality on CAMEO data set for ProQ2 (top) and ProQ3 (bottom) trained
  with a SVM (left)   or Theano (right). RMSE is a root mean square error between the prediction and the target value (lower values are better). (a) ProQ2, SVM, (b) ProQ2D, Theano, (c) ProQ3, SVM, (d) ProQ3D, Theano}
\label{fig:local_cameo}
\end{figure}

Additionally, we provide the local prediction (residue-level) scatter
plots for CASP and CAMEO data sets
(Figure~\ref{fig:local_casp11}~and~\ref{fig:local_cameo}). The density
plots suggest that the deep learning versions of methods are better at
distinguishing low quality models on CASP11 data set (S-score $< 0.1$)
and high quality models on CAMEO data set (S-score $> 0.9$).

In the benchmark we used the updated ProQ version (ProQres) that is able to
predict the local model quality~\citep{Wallner16522791}. We have also
compared our methods with two other state-of-the-art methods:
Qmean~\citep{Benkert17932912} and Qprob~\citep{Cao27041353}, as well as a reference consensus method Pcons~\citep{Lundstrom11604541}.

\section{Methods}
\subsection{Architecture}

The chosen architecture is that of a Multi Layer Perceptron (MLP)
\citep{mlp} with two hidden layers of 600 and 200 hidden units
respectively, and using ReLU \citep{relu} as a non linearity.  The
parameters are optimised with Adadelta \citep{adadelta}, a
modification on the Stochastic Gradient Descent that can automatically
adapt the learning rate, trying to optimise the mean squared error.
We used two regularisation methods: dropout \citep{dropout} and $L^2$
penalty on the weights.
Varying the number of neurons in each layer did not yield a significant difference in the results.

Adding more hidden layers gives worse results because the convergence is slowed down \citep{difficult_deep}, and the model does not gain expressiveness since it is already an universal approximator \citep{universal_approximator}.

\subsection{Dropout}
With dropout, at training time, we set to zero ("drops") a random
fraction $p$ of the hidden neurons of each layer in each iteration,
while multiplying the rest by a factor $1/p$.  For a model with $H$
hidden units, the output at test time is the geometric mean of the
$2^H$ possible combinations, each of which has seen at most one data
point, and are strongly regularised towards the consensus.
Following the original paper, we set $p=0.5$.

\subsection{$L^2$ penalty}
With the aim to discourage large weights and improve generalisation accuracy, we add a term to the loss function:
\begin{equation}
\lambda \sum W_{ij}^2,
\end{equation}
where $\lambda$ is a scalar and $W$ are the weight matrices of the MLP.

This prevents weights to grow too large unless they are sufficiently
beneficial, that is, they can overcome the penalty in the loss
function.

We tested different training penalties ranging from $10^{-4}$ to $10^{-12}$, with minor effects on the quality of the results.

\subsection{Implementation}
The implementation was done using the Keras framework \citep{keras}
and the Theano backend \citep{theano}, and trained on GPUs. However,
the ProQ3D program does not require GPU to run, because prediction
works fast enough on CPU (less than 1 second per model, excluding
feature generation).

\clearpage

\begin{table}[!ht]
\caption{
{\bf Performance of different QA methods on CAMEO data set }}
\centering
\begin{tabular}{p{1.2cm}|p{1.2cm}p{1.2cm}p{1.2cm}p{1.2cm}p{1.2cm}} 
	& CC-glob & CC-target & CC-loc & CC-model & GDT\_loss \\
\hline	
  ProQ & 0.67 & 0.54 & 0.52 & 0.45 & 0.04 \\
  Qprob  & 0.68 & 0.53 & - & - & 0.04 \\
  Qmean  & 0.68 & 0.48 & 0.54 & 0.47 & 0.05 \\
  ProQ2 & 0.75 & 0.57 & 0.60 & 0.50 & 0.04 \\
  ProQ3 & 0.79 & 0.58 & 0.65 & 0.55 & 0.04 \\
  ProQ2D & 0.79 & 0.60 & 0.64 & 0.52 & 0.04\\
  ProQ3D  & 0.82 & 0.60 & 0.69 & 0.58 & 0.04 \\
\hline
 & \multicolumn{5}{c}{Consensus based methods} \\
\hline
  Pcons  & 0.89 & 0.67 & 0.81 & 0.70 & 0.05 \\
\end{tabular}
\label{tab:correlations}
\end{table}

\bibliographystyle{natbib}